\documentclass[12pt,preprint]{aastex}
\usepackage{epsf,graphics}
\newcommand{\simgt}{\lower.5ex\hbox{$\; \buildrel > \over \sim \;$}}
\newcommand{\simlt}{\lower.5ex\hbox{$\; \buildrel < \over \sim \;$}}

\shorttitle{Mean-pairwise peculiar velocity}
\shortauthors{Fukushige \& Suto}
\begin{document}

\title{Mean-pairwise peculiar velocity in cosmological N-body
simulations: time-variation, scale-dependence and stable condition}

\author{Toshiyuki Fukushige}
\affil{Department of General Systems Studies, College of Arts and
Sciences \\ University of Tokyo, 3-8-1 Komaba, Meguro-ku, 
Tokyo 153-8902, Japan.}
\email{fukushig@provence.c.u-tokyo.ac.jp}

\and

\author{Yasushi Suto} 
\affil{Department of Physics and Research Center for the Early Universe
(RESCEU)\\ School of Science, University of Tokyo, Tokyo 113-0033,
Japan.}
\email{suto@phys.s.u-tokyo.ac.jp}

\received{2001 April 18}
\accepted{2001 June 29}

%-----------------------------------------------------------------------
\begin{abstract} 
We report on the detailed analysis of the mean-pairwise peculiar
velocity profile in high-resolution cosmological N-body simulations
($N=8.8\times 10^6$ particles in a sphere of $50 \sim 200$Mpc radius).
In particular we examine the validity and limitations of the stable
condition which states that the mean {\it physical} separation of
particle pairs is constant on small scales.  We find a significant
time-variation (irregular oscillatory behavior) of the mean-pairwise
peculiar velocity in nonlinear regimes.  We argue that this behavior is
not due to any numerical artifact, but a natural consequence of the
continuous merging processes in the hierarchical clustering universe.
While such a time-variation is significant in a relatively local patch
of the universe, the global average over a huge spatial volume
$\simgt$(200Mpc)$^3$ does not reveal any {\it systematic} departure from
the stable condition.  Thus we conclude that the mean-pairwise peculiar
velocity is rather unstable statistics but still satisfies the stable
condition when averaged over the cosmological volume.
\end{abstract}

%-----------------------------------------------------------------------

%%%%%%%%%%%%%%%%%%%%%%%%%%%%%%%%%%%%%%%%%%%%%%%%%%%%%%%%%%%%%%%%%%%%%%%%%

\section{Introduction}

Hubble's law indicates that galaxies at cosmological distances are
approximately at rest with respect to the comoving frame of the
universe.  Dynamics of self-gravitating objects on much smaller scales,
on the other hand, is significantly more complex and definitely deviates
from the global cosmic expansion.  A simple and natural question is
under what conditions the nonlinear self-gravitating systems completely
decouple from the expansion of the universe.  If one considers an
isolated system strictly in a virial equilibrium, one can show that the
physical separation of any particle-pair in the system does not change
on average.  This is supposed to be a good approximation, for instance,
for the separation between the Earth and the Sun.

In a hierarchical clustering universe, however, no system can remain
isolated in a strict sense, and the virial equilibrium may not be
realized either.  The condition that the mean separation $r$ of galaxy
pairs is constant in physical coordinates is translated as
%%%%%%%%%%%%%%%%%%%%%%%%%%%%%%%%%%%%%%%%%%%%%%%%%%%%%%%%%%%%%%%%%%%%%
\begin{equation}
v_{12}(r,z) + H(z)r = 0 ,
\end{equation}
%%%%%%%%%%%%%%%%%%%%%%%%%%%%%%%%%%%%%%%%%%%%%%%%%%%%%%%%%%%%%%%%%%%%%
where $v_{12}(r,z)$ is the mean-pairwise peculiar velocity at that scale
and $H(z)$ is the Hubble parameter at the redshift $z$.  Davis \&
Peebles (1977) showed that the two-point correlation function in
nonlinear regimes approaches the stable clustering solution, $\xi(r)
\propto r^{-3(n+3)/(n+5)}$ if the above {\it stable condition} is exact
and the initial density fluctuations follow the scale-free power
spectrum $P(k) \propto k^n$.  As in this example, the stable condition
plays an important role in modeling the cosmological nonlinear power
spectrum, and in fact this idea has been used in more accurate
predictions (Hamilton et al.  1991; Nityananda \& Padmanabhan 1994;
Jain, Mo, \& White 1995; Peacock \& Dodds 1996; Suto \& Jing 1997;
Suginohara et al.  2001).

This stable condition has been tested against cosmological N-body
simulations by various authors (e.g., Efstathiou et al.  1988;
Suginohara et al.1991; Suto 1993).  In particular Jain (1997) discussed
in detail the departure from the stable condition using his P$^3$M
simulations with $(100^3 \sim 144^3)$ particles.  We note here that he
mostly estimated the mean-pairwise peculiar velocity profile {\it
indirectly} using the evolution of the volume-averaged two-point
correlation function $\bar\xi(r)$ and the pair-conservation equation:
%%%%%%%%%%%%%%%%%%%%%%%%%%%%%%%%%%%%%%%%%%%%%%%%%%%%%%%%%%%%%%%%%%%%%%%
\begin{equation}
\label{eq:meanvelocity}
{v_{12}(r,z) \over H r}
= {1 \over 3[1+\xi(r,z)]} {\partial \bar\xi(r,z) \over \partial \ln(1+z)}
\end{equation}
%%%%%%%%%%%%%%%%%%%%%%%%%%%%%%%%%%%%%%%%%%%%%%%%%%%%%%%%%%%%%%%%%%%%%%%
because he found the data of $v_{12}(r)$ are much noisier than those of
$\xi(r)$. Caldwell et al.  (2001) elaborated the findings of Jain (1997)
and proposed a universal function of $v_{12}(r)/H r$ in terms of
$\bar\xi(r)$.  Jing (2001), on the other hand, found that the isolated
virialized halos yield $v_{12}(r)/H r \sim -1$, i.e., the stable
condition holds for those halos.  Those simulations, however, did not
yet address, in a convincing manner, the validity of the {\it
cosmologically averaged} stable condition in strongly nonlinear regimes,
because of the indirect method of evaluation and/or uncertainties due to
the poor statistics.  Actually the stable condition may be regarded as a
mere assumption at this point, and several analytic arguments against
the validity of the stable condition have been proposed (Kanekar 2000;
Yano \& Gouda 2000; Ma \& Fry 2001).

Therefore we attempt to investigate the validity and limitations of the
stable condition without any simplifying assumptions as possible using
the cosmological N-body simulation. We perform simulations in a fairly
different and complementary fashion compared with the previous ones.
Specifically we adopt physical (rather than comoving) coordinates so
that the departure from the stable condition is detected more clearly, a
different gravity solver based on the hierarchical tree algorithm
(Barnes \& Hut 1986), and a spherical vacuum (rather than periodic)
boundary condition.  Also we pay particular attention to the temporal
and spatial variation of the mean-pairwise velocity profile, which has
not been considered before.

%%%%%%%%%%%%%%%%%%%%%%%%%%%%%%%%%%%%%%%%%%%%%%%%%%%%%%%%%%%%%%%%%%%%%
\section{Simulation method}

Cosmological N-body simulations are usually performed in comoving
coordinates since the deviation from the Hubble expansion is of main
interest in gravitational clustering, especially in linear to mildly
nonlinear regimes. The stable condition, however, states that the
particle pair separation is constant in physical coordinates, and thus
we decided to integrate the system using these coordinates (see also
Fukushige \& Makino 1997, 2001). More specifically, we evolve the system
according to the following equation of motion:
%%%%%%%%%%%%%%%%%%%%%%%%%%%%%%%%%%%%%%%%%%%%%%%%%%%%%%%%%%%%%%%%%%%%%%
\begin{equation}
\label{eq:eom}
{d^2 {\bf r}_i \over dt^2} 
= \displaystyle{\sum_j
{Gm_j({\bf r}_j - {\bf r}_i) 
\over (|{\bf r}_j - {\bf r}_i|^2 + \varepsilon_{\rm grav}^2)^{3/2} 
}} 
+ H_0^2\lambda_0{\bf r}_i , 
\end{equation}
%%%%%%%%%%%%%%%%%%%%%%%%%%%%%%%%%%%%%%%%%%%%%%%%%%%%%%%%%%%%%%%%%%%%%%
where ${\bf r}_i$ and $m_i$ are the physical coordinate and mass of the
$i$-th particle, and $H_0 = 100 h \, {\rm km}\cdot{\rm s}\cdot{\rm
Mpc}^{-1}$ and $\lambda_0$ denote the Hubble constant and the
dimensionless cosmological constant at the present epoch.  The second
term in the right-hand-side expresses an acceleration with respect to
the center of the simulation sphere (see below) in the presence of the
non-vanishing cosmological constant. We use the constant gravitational
softening length $\varepsilon_{\rm grav}=5$kpc in {\it physical}
coordinates.

We construct the initial condition for each simulation run as follows;
first, we distribute $N=256^3$ equal-mass particles in a cube of $8R^3$
at a redshift $z=z_i$ using the initial condition generator in the Hydra
code (Couchman et al.  1995).  Then we extract a sphere of radius $R$
from the cube, and thus $\sim 8.8\times 10^6$ particles are left in the
simulation sphere.  While we neglect the external tidal field outside
the sphere, we made sure that the resulting two-point correlation
functions are in good agreement with the Peacock \& Dodds prediction on
scales below $\sim 10$Mpc, and thus our results at those scales are not
affected by the assumed boundary condition.  Finally, we add the Hubble
flow for each particle, $H(z_i){\bf r}_i$, where
%%%%%%%%%%%%%%%%%%%%%%%%%%%%%%%%%%%%%%%%%%%%%%%%%%%%%%%%%%%%%%%%%%%%
\begin{equation}
H(z) = H_0 \sqrt{\Omega_0 (1+z)^3
       +(1-\Omega_0-\lambda_0)(1+z)^2 + \lambda_0 } , 
\end{equation}
%%%%%%%%%%%%%%%%%%%%%%%%%%%%%%%%%%%%%%%%%%%%%%%%%%%%%%%%%%%%%%%%%%%%
and $\Omega_0$ is the density parameter at present.  We consider six
cosmological models listed in Table \ref{tab:parameters}; Standard, Open
and Lambda Cold Dark Matter models (SCDM, SCDM50, SCDM200, OCDM, and
LCDM) and a Poisson model in the Einstein - de Sitter universe (EdS0).
The SCDM50 and SCDM200 models adopt $R=50$ Mpc and $200$ Mpc,
respectively, while the other four models use $R=100$ Mpc.  The
amplitudes of the power spectrum in CDM models are normalized using the
top-hat filtered mass variance at $8 h^{-1}$ Mpc according to the
cluster abundance (Kitayama \& Suto 1997), and the EdS0 model adopts the
same normalization of SCDM for reference.

We integrate equation (\ref{eq:eom}) using a leap-flog integrator with
shared and constant timestep, $\Delta t = (t_0-t_i)/1024$ for the
SCDM200 model and $\Delta t = (t_0-t_i)/2048$ for other models, where
$t_0$ and $t_i$ denote the present and initial cosmic times,
respectively (strictly speaking, we use $\Delta t = (t_0-t_i)/16384$
for $z>14.4$ in EdS0 case only).  The force calculation is made with
the Barnes-Hut tree code (Barnes \& Hut 1986) on GRAPE-5 (Kawai et al.
2000), a special-purpose computer designed to accelerate $N$-body
simulations.  Our code adopts the Barnes modified tree algorithm
(Barnes 1990) which is implemented on the GRAPE systems by Makino
(1991); see Kawai et al. (2000) for details. In order to maintain the
accuracy of the force calculation for the present purpose, we use a
rather small value for the opening parameter $\theta = 0.4$.  The
simulations presented below need $\sim 150$ secs per timestep, and
thus one run (2048 timesteps) is completed in 85 CPU hours with a
GRAPE-5 board on a host workstation (21264 Alpha chip) at the
Astronomical Data Analysis Center of the National Astronomical
Observatory, Japan.

%%%%%%%%%%%%%%%%%%%%%%%%%%%%%%%%%%%%%%%%%%%%%%%%%%%%%%%%%%%%%%%%%%%%%
\section{Mean-pairwise peculiar velocity}

\subsection{Scale-dependence and the stable condition}

We compute the mean-pairwise peculiar velocity, $v_{12}(r)$, by
averaging over particle pairs within $0.8R$ from the center of the
simulation sphere; for particle separation $r<1.0/(1+z)$ Mpc we use all
pairs in the entire sphere, while for $r>1.0/(1+z)$ Mpc we first
randomly select $0.02\%$ center particles, find all pairs for those
particles, and then average their pairwise peculiar velocities.

\begin{figure}[b]
\begin{center}
\leavevmode\epsfxsize=14.0cm \epsfbox{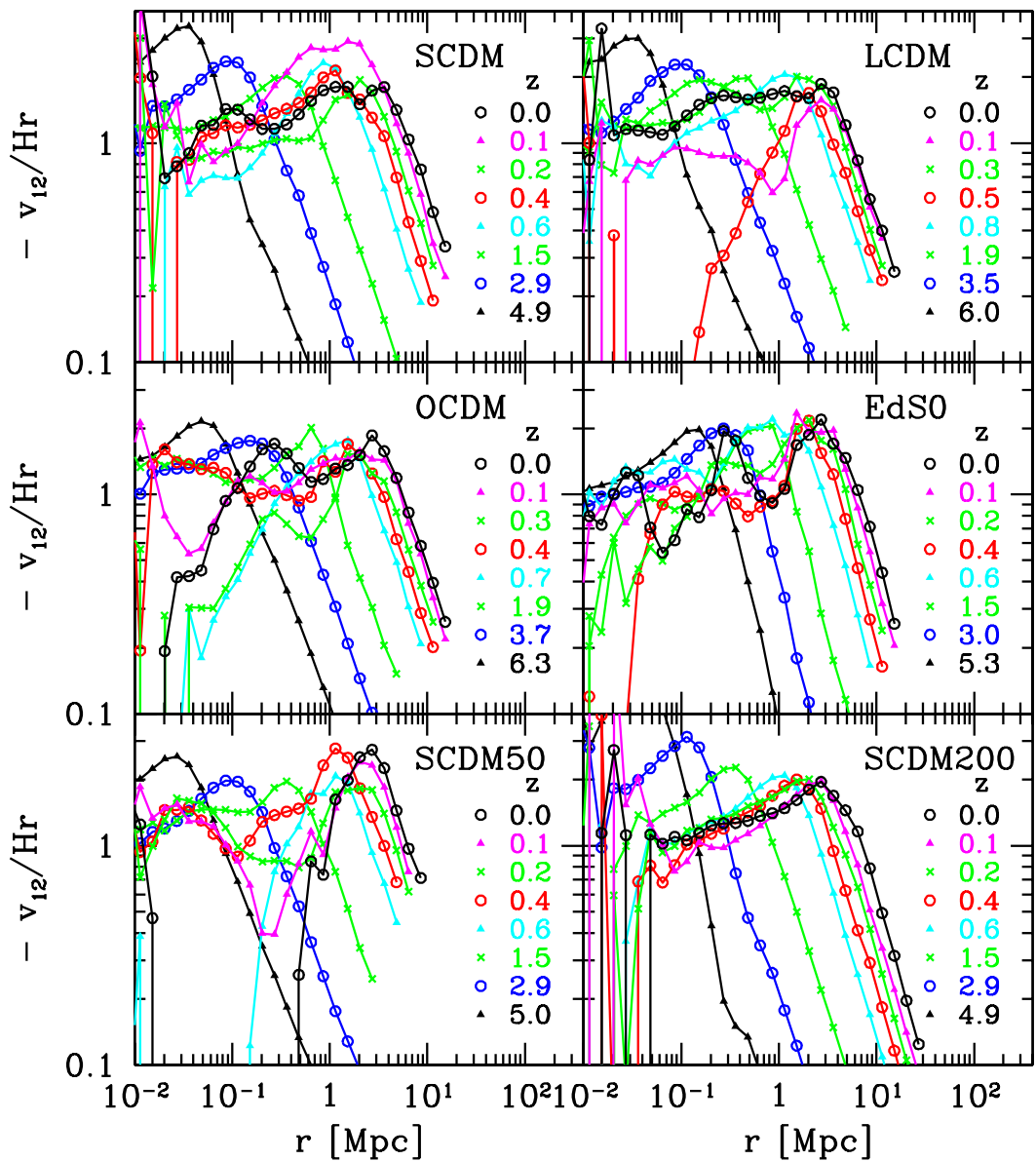}
% \leavevmode\epsfxsize=14.0cm \epsfbox{fs_fig1_cmyk.eps}
\end{center} 
  \figcaption{The normalized mean-pairwise peculiar velocity,
$-v_{12}/Hr$ as a function of the pair separation $r$ at different
redshifts.  \label{fig:v12r}}
\end{figure}

Figure~\ref{fig:v12r} shows the scale-dependence of the {\it normalized}
mean-pairwise peculiar velocity, $-v_{12}(r)/Hr$, at different redshifts
($z \simlt 6$). In linear regimes, the peculiar velocity of pairs
$v_{12}(r)$ is negative but smaller than the Hubble velocity $Hr$, and
thus the pair-separation still increases in physical length.  As
clustering proceeds, a given object starts collapsing which corresponds
to $- v_{12}(r)/Hr>1$. After experiencing this collapse phase, the
system reaches quasi-equilibrium close to $-v_{12}(r)/Hr=1$.  As
Figure~\ref{fig:v12r} illustrates, the ratio seems to approach
$-v_{12}(r)/Hr=1$ (the stable condition) at $r \rightarrow 0$ in all
models, but only with significant scatter and variations.

\begin{figure}[htbp]
\begin{center}
\leavevmode\epsfxsize=14.0cm \epsfbox{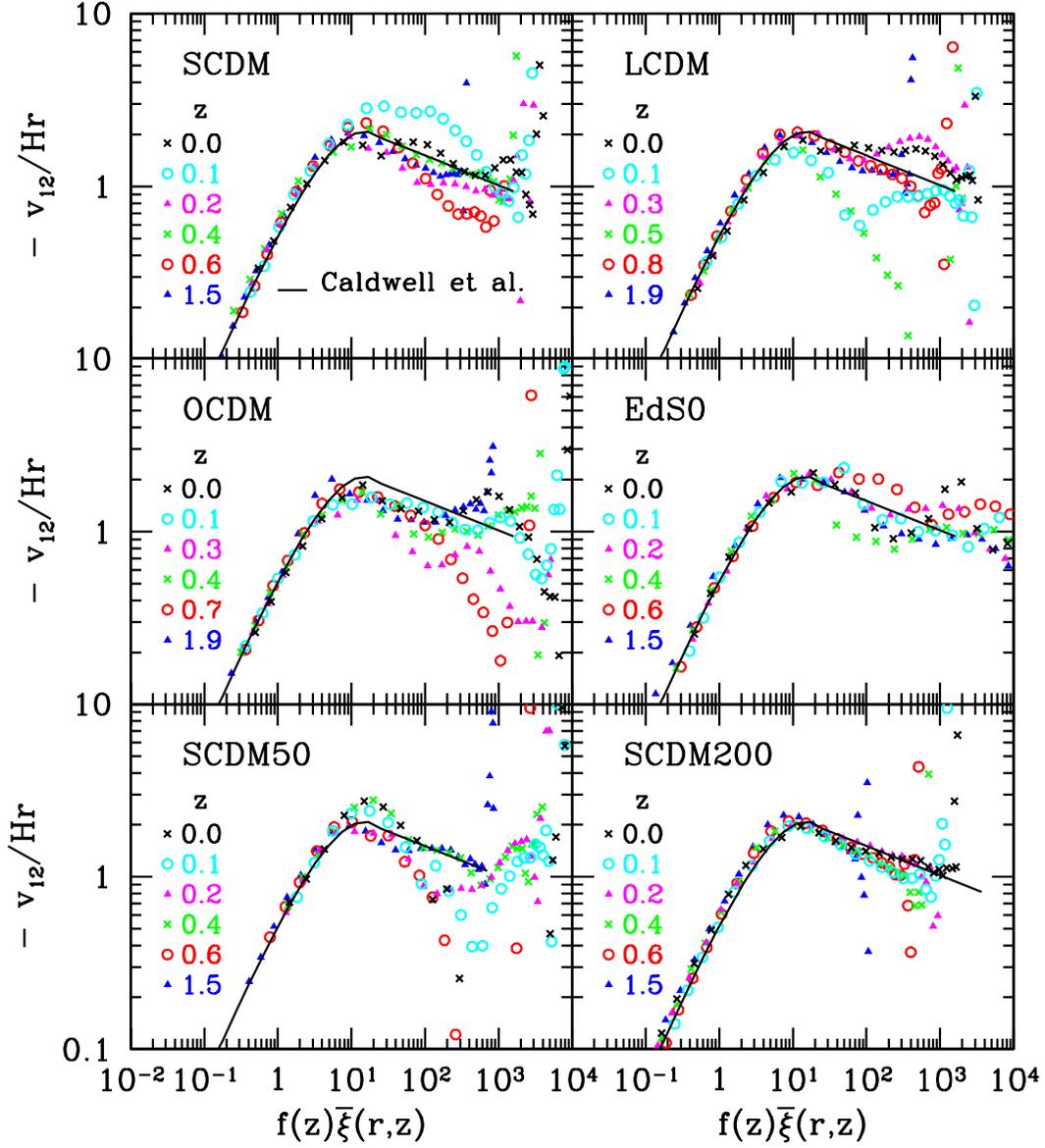}
% \leavevmode\epsfxsize=14.0cm \epsfbox{fs_fig2_cmyk.eps}
\end{center} 
  \figcaption{The normalized mean-pairwise peculiar velocity,
$-v_{12}/Hr$ as a function of $f(z)\bar{\xi}(r;z)$ at different
redshifts. Sold lines indicate the fitting formula proposed by Caldwell
 et al. (2001). \label{fig:v12fxi}}
\end{figure}

Figure~\ref{fig:v12fxi} replots the same data of Figure~\ref{fig:v12r}
as a function of $f(z)\bar\xi(r,z)$ at different epochs ($z \simlt 2$),
following the scaling of Caldwell et al.  (2001), where $f(z)$ is the
logarithmic derivative of the linear growth rate with respect to the
cosmic scale factor.  The solid lines in the figure indicate the fitting
formula of Caldwell et al.  (2001).  In the linear and quasi-nonlinear
regimes where $f(z)\bar\xi(r,z) \simlt 10$, our results are in good
agreement with the scaling.  In more strongly nonlinear regimes,
however, the ratio $-v_{12}(r)/Hr$ varies significantly as a function of
$r$ and $z$, and especially shows somewhat irregular oscillatory
behavior in time.  In those regimes, our results marginally reproduce
the fitting formula of Caldwell et al.  (2001) only after averaging over
time and/or ensembles.  Even then the small but systematic deviation
from the fitting formula is visible, especially in OCDM model, which
might be explained by the difference of the method of estimating
$-v_{12}(r)/Hr$; Figure 2 in Jain (1997) shows the similar systematic
deviation between the direct estimate and that computed using the
pair-conservation equation.  In fact, the degree of those variations
sensitively depends on the size of the simulation volume
itself. Comparison of the SCDM models with $R=50$, 100, and 200 Mpc
shows that the variation becomes systematically smaller because of the
statistical average over the larger volume.  On the other hand, the
spatial resolution of the SCDM200 model is not so good as SCDM50, and
cannot probe the strongly nonlinear scales corresponding to
$f(z)\bar\xi(r,z) \simgt 1000$.

For the same reason, Caldwell et al.  (2001) remarked that their formula
is valid for $f(z)\bar\xi(r,z) \simlt 1000$, and that it is not clear
whether $-v_{12}(r)/Hr$ converges to unity or possibly oscillates beyond
the value. Our result in Figure~\ref{fig:v12fxi} indicates that the
ratio shows an appreciable oscillation even for $f(z)\bar\xi(r,z) \simgt
100$. Nevertheless its time-average in all models is {\it not
inconsistent} with the stable condition asymptotically, i.e., at
$f(z)\bar\xi(r,z) = (1000 \sim 10000)$, and we interpret this to suggest
that the stable condition is satisfied after averaging over the
cosmological volume.

\subsection{Time-variation}

\begin{figure}[htbp]
\begin{center}
\leavevmode\epsfxsize=14.0cm \epsfbox{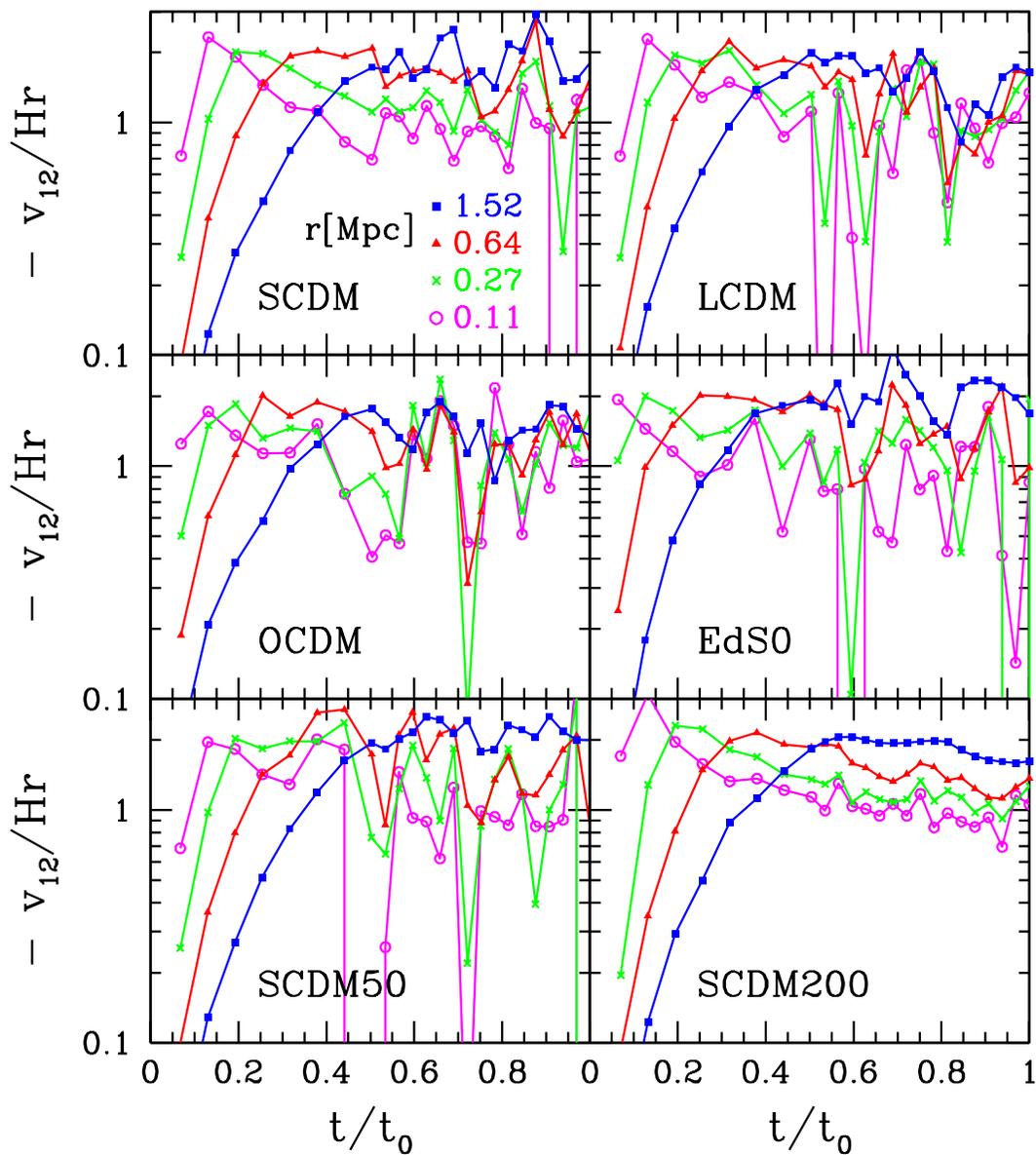}
% \leavevmode\epsfxsize=14.0cm \epsfbox{fs_fig3_cmyk.eps}
\end{center} 
  \figcaption{Evolution of $-v_{12}/Hr$ at fixed pair separations
 $r=0.11$, 0.27, 0.64 and 1.52 Mpc in physical coordinates against the
 cosmic time.
 \label{fig:evolv12}}
\end{figure}

In order to understand the origin of the significant variation of
$-v_{12}(r)/Hr$, we plot the ratio at fixed physical separations against
the cosmic time in units of the present value $t_0$
(Fig.\ref{fig:evolv12}).  As the clustering in the corresponding scale
proceeds, $-v_{12}(r)/Hr$ monotonically increases in early epochs, but
once the separation becomes below the typical collapse scale
corresponding to $f(z)\bar\xi(r,z) \sim 10$, $-v_{12}(r)/Hr$ exceeds
unity and then starts sporadic oscillations.

Jain (1997) also noticed that the direct estimate of the mean-pair wise
velocity is quite noisy in his simulations adopting very different
gravity solver, integration scheme and the boundary condition from our
present ones.  We still suspect that the numerical errors due to the
two-body relaxation and the integration scheme are not negligible at
scales much less than $0.1$Mpc where the Hubble flow amounts to $\sim
10$km/s while the peculiar velocity dispersion exceeds $\sim 500$km/s.
Therefore our results below those scales, corresponding to
$f(z)\bar\xi(r,z) > 10^3$ at $z=0$, may be interpreted with
caution. Nevertheless the significant time-variation is visible even for
$f(z)\bar\xi(r,z) > 10$. Those facts mentioned above suggest that the
significant time-variation is not due to any numerical artifact.

We ran several other models with different power spectra, number of
particles, and simulation boxsizes, and found that the time-variation
persists in all cases, and even becomes stronger for smaller volume
runs.  This is clearly exhibited for the SCDM50 model in Figure
\ref{fig:evolv12}.  We interpret this behavior as a result of the
continuous merging of small-scale objects in the hierarchical clustering
universe.  Once an over-dense region decouples from the cosmic expansion
and becomes self-gravitating, it tends to approach a state of the virial
equilibrium, at least temporarily.  Such a system, however, is rarely
isolated, but rather forms a part an over-dense region on larger scales.
Therefore it subsequently experiences disruption and moves to another
state of the virial equilibrium corresponding to the higher dynamical
temperature in the course of hierarchical merging.  Such dynamical
behavior should exhibit s significant time-variation if traced at a
fixed physical separation. Thus the true mean-pairwise velocity can be
estimated only by averaging over a huge cosmological volume; our results
imply that the average over $\sim (100$Mpc$)^3$ is not still robust and
that at least $\simgt (200$Mpc$)^3$ is required to have a reliable
estimate for $v_{12}(r)$ even on scales below 1Mpc.

\section{Conclusions and discussion}

Using a series of high-resolution $N$-body simulations performed in
physical coordinates, we compute the mean-pairwise peculiar velocity
profile directly from the particle data. In linear and quasi-nonlinear
regimes, we confirmed that our simulation data obey the scaling and the
fitting formula proposed by Caldwell et al.  (2001).  In the highly
nonlinear regime where the stable condition is conventionally assumed,
we found a significant time-variation in the normalized mean-pairwise
peculiar velocity $-v_{12}(r)/Hr$, i.e., {\it the stable condition for
the physical pair-separation is not stable in time}.  This behavior can
be understood as a natural consequence of the continuous merging
processes in the hierarchical clustering universe.  This variation can
be reduced only by the ensemble average over the cosmological volume
$\simgt (200 {\rm Mpc})^3$. This is in a sense surprising since the
two-point correlation function is a fairly stable statistics and their
sample-to-sample variation is not so big \citep{itoh92}.  On the other
hand, our results do not exhibit any clear signal for the {\it
systematic} departure from the stable condition given the large
time-variation.  Thus we conclude that while the mean-pairwise peculiar
velocity is rather unstable statistics, the stable condition is still
valid after the averaging over the entire volume of the universe.

Combined with the fact that the stable condition seems to be satisfied
in individual {\it isolated} halos (Jing 2001), our results rule out
some recent claims that $-v_{12}(r)/Hr$ approaches an asymptotic value
smaller than $1/2$ (Kanekar 2000), but may not be inconsistent with the
weaker departure predicted by Ma \& Fry (2000).  Considering the
intrinsic nature of the origin for the time-variation, a more definitive
answer for the validity of the stable condition for cosmologically
averaged pairs requires numerical simulations which have both the higher
mass-resolution and the larger simulation volume than our current runs.

\acknowledgements

We thank Y.P.Jing, A.Kawai, J.Makino, T.Suginohara, and A.Taruya for
useful discussions, and an anonymous referee for pointing out the
importance of the sample-to-sample variation.  We gratefully acknowledge
the use of the initial condition generator in the publicly available
code {\it Hydra} developed by H.M.P.Couchman, P.A.Thomas, and
F.R.Pearce.  Numerical computations were carried out on the GRAPE system
at ADAC (the Astronomical Data Analysis Center) of the National
Astronomical Observatory, Japan.  This research was supported in part by
the Grant-in-Aid by the Ministry of Education, Science, Sports and
Culture of Japan (07CE2002, 12640231), and by the Research for the
Future Program of Japan Society for the Promotion of Science (JSPS-RFTP
97P01102).

\bigskip
\bigskip

%%%%%%%%%%%%%%%%%%%%%%%%%%%%%%%%%%%%%%%%%%%%%%%%%%%%%%%%%%%%%%%%%%%%%%

%%%%%%%%%%%%%%%%%%%%%%%%%%%%%%%%%%%%%%%%%%%%%%%%%%%%%%%%%%%%%%%%%%%%%%%

\clearpage

%%%%%%%%%%%%%%%%%%%%%%%%%%%%%%%%%%%%%%%%%%%%%%%%%%%%%%%%%%%%%%%%%%%%%
\begin{deluxetable}{cccccccc}
\footnotesize
\tablecaption{Simulation parameters
\label{tab:parameters}
}
\tablewidth{0pt}
\tablehead{
\colhead{Model} & \colhead{$\Omega_0$}& \colhead{$\lambda_0$} & 
\colhead{$h$} & \colhead{$P(k)$} & \colhead{amplitude}  &  
\colhead{$1+z_i$} & \colhead{$m$($M_{\odot}$)} 
}
\startdata
SCDM & 1.0 & 0.0 & 0.5 & $\Gamma=0.5$ & $\sigma(8h^{-1}{\rm Mpc})=0.6$ & 
25.0 & $3.3\times 10^{10}$ \\
LCDM & 0.3 & 0.7 & 0.7 & $\Gamma=0.21$ & $\sigma(8h^{-1}{\rm Mpc})=1.0$ & 
33.3 & $1.9 \times 10^{10}$\\
OCDM & 0.3 & 0.0 & 0.7 & $\Gamma=0.21$ & $\sigma(8h^{-1}{\rm Mpc})=1.0$ & 
33.3 & $1.9 \times 10^{10}$ \\
EdS0 & 1.0 & 0.0 & 0.5 & $n=0$ & $\sigma(8h^{-1}{\rm Mpc})=0.6$ & 
100.0 & $3.3\times 10^{10}$ \\
SCDM50 & 1.0 & 0.0 & 0.5 & $\Gamma=0.5$ & $\sigma(8h^{-1}{\rm Mpc})=0.6$ & 
30.0 & $4.1\times 10^{9}$ \\
SCDM200 & 1.0 & 0.0 & 0.5 & $\Gamma=0.5$ & $\sigma(8h^{-1}{\rm Mpc})=0.6$ & 
25.0 & $2.6\times 10^{11}$ \\
\enddata
\end{deluxetable}
%%%%%%%%%%%%%%%%%%%%%%%%%%%%%%%%%%%%%%%%%%%%%%%%%%%%%%%%%%%%%%%%%%%%%

\clearpage

%%%%%%%%%%%%%%%%%%%%%%%%%%%%%%%%%%%%%%%%%%%%%%%%%%%%%%%%%%%%%%%
%%%%%%%%%%%%%%%%%%%%%%%%%%%%%%%%%%%%%%%%%%%%%%%%%%%%%%%%%%%%%%%%

%%%%%%%%%%%%%%%%%%%%%%%%%%%%%%%%%%%%%%%%%%%%%%%%%%%%%%%%%%%%%%%
%%%%%%%%%%%%%%%%%%%%%%%%%%%%%%%%%%%%%%%%%%%%%%%%%%%%%%%%%%%%%%%%

%%%%%%%%%%%%%%%%%%%%%%%%%%%%%%%%%%%%%%%%%%%%%%%%%%%%%%%%%%%%%%%
%%%%%%%%%%%%%%%%%%%%%%%%%%%%%%%%%%%%%%%%%%%%%%%%%%%%%%%%%%%%%%%%

\end{document}